\documentclass[journal]{IEEEtran}  

\usepackage{graphics} 
\usepackage{epsfig} 
\usepackage{amsmath,amsfonts,amssymb,amsthm,array}  

\usepackage[inline]{enumitem}

\newtheorem{remark}{Remark}

\newcommand*{\mc}[1]{\mathcal{#1}}
\newcommand*{\mbb}[1]{\mathbb{#1}}
\newcommand*{\mb}[1]{\boldsymbol{#1}}
\newcommand*{\norm}[1]{\left\lVert #1 \right\rVert}
\DeclareMathOperator{\atantwo}{atan2}

\usepackage{algpseudocode}
\usepackage{subcaption}
\usepackage{outlines}
\usepackage{booktabs}

\usepackage{flushend}

\usepackage{color}

\begin{document}

\title{
Intelligent Knowledge Distribution: Constrained-Action POMDPs for\\Resource-Aware Multi-Agent Communication
}

\author{Michael C. Fowler$^{1}$, T. Charles Clancy$^{2}$, and Ryan K. Williams$^{3}$
  \thanks{$^{1}$Michael C. Fowler is a research scientist with the Hume Center for National Security and Technology, and a Ph.D. Candidate in Electrical and Computer Engineering,
          Virginia Tech, Blacksburg, VA 24060, USA,
          {\tt\small mifowler@vt.edu}}%
  \thanks{$^{2}$T. Charles Clancy is with the Department of Electrical Engineering, Virginia Tech,
          Blacksburg, VA 24060, USA,
          {\tt\small tcc@vt.edu}}%
  \thanks{$^{3}$Ryan K. Williams is with the Department of Electrical and Computer Engineering, Virginia Tech,
           Blacksburg, VA 24060, USA,
          {\tt\small rywilli1@vt.edu}}%
}


\maketitle

\begin{abstract}
  This paper addresses a fundamental question of multi-agent knowledge distribution: what information should be sent to whom and when, with the limited resources available to each agent?
  Communication requirements for multi-agent systems can be rather high when an accurate picture of the environment and the state of other agents must be maintained.
  To reduce the impact of multi-agent coordination on networked systems, e.g., power and bandwidth, this paper introduces two concepts for partially observable Markov decision processes (POMDPs):
  \begin{enumerate*}
    \item \emph{action-based constraints} which yield constrained-action POMDPs (CA-POMDPs); and
    \item \emph{soft probabilistic constraint satisfaction} for the resulting infinite-horizon controllers.
  \end{enumerate*}
  To enable constraint analysis over an infinite horizon, an unconstrained policy is first represented as a Finite State Controller (FSC) and optimized with policy iteration.
  The FSC representation then allows for a combination of Markov chain Monte Carlo and discrete optimization to improve the probabilistic constraint satisfaction of the controller while minimizing the impact to the value function.
  Within the CA-POMDP framework we then propose \emph{Intelligent Knowledge Distribution (IKD)} which yields per-agent policies for distributing knowledge between agents subject to interaction constraints.
  Finally, the CA-POMDP and IKD concepts are validated using an asset tracking problem where multiple unmanned aerial vehicles (UAVs) with heterogeneous sensors collaborate to localize a ground asset to assist in avoiding unseen obstacles in a disaster area.
  The IKD model was able to maintain asset tracking through multi-agent communications while only violating soft power and bandwidth constraints 3\% of the time, while greedy and naive approaches violated constraints more than 60\% of the time.
\end{abstract}

\IEEEpeerreviewmaketitle

\section{Introduction}
\IEEEPARstart{D}{ecentralized} coordination of autonomous systems in the field requires a delicate balance between system objectives and the burden on limited resources such as power, bandwidth, computation, etc.
The resilience of decentralized approaches is clear, allowing agents to act independently with varying levels of information about other agents' states, observations, and actions.
However, decentralized systems by their nature must \emph{intelligently} consume resources to remain feasible.
In this paper, we propose a framework for \emph{Intelligent Knowledge Distribution} (IKD) which decides on what information is transmitted to whom and when, while constraining the impact coordination has on the limited resources available to each agent.

In disaster response or military operations, remote sensing by autonomous systems provides a stream of data back to critical personnel.
Disaster recovery and other active sensing regimes often rely on high bandwidth links that are stretched to their limits under unpredictable channel and link qualities in support of electrooptical, chemical sensors, electronic sensing, etc. \cite{tuna2014unmanned}.
At the same time, the communication necessary for multiple agents to coordinate needs to stay within constrained allowances as to not impact the flow of critical information in the system.
Along with typical bandwidth constraints, unmanned aerial vehicles (UAVs) for example have the additional limitation of battery power that needs to be preserved to maximize their flight time in support of field operations.

With the above concerns in mind, in this work we assume each agent is executing its own decision-making in a decentralized network, i.e., our underlying model is a decentralized partially observable Markov decision process (Dec-POMDP) \cite{omidshafiei2015dec,oliehoek2016concise,seuken2012dpdec}.  In the Dec-POMDP setting, each agent has local state observations but only indirect observations of the environment and other agents.
Thus, an agent can run independently without the necessity of communicating or having information about other agents' states or rewards, but this will lead to sub-optimal behavior.
Indeed, the key problem is to determine from local observations what information neighboring agents will need and when they will need it, improving the overall effectiveness of a coordination objective.
We propose methods to achieve these goals while respecting the resource constraints of fielded systems, eliminating the common assumption that communications are instantaneous and free \cite{capitan2013}.
In particular, we aim to abstract communication decisions away from other joint agent policies (e.g., for motion, tasking, etc.) to construct an agnostic ``plug-and-play'' capability to mitigate combinatorial explosion and provide an approach that may utilize information from other models similar to Concurrent MDPs \cite{girard2015}.

In this paper, the Constrained-Action POMDP (CA-POMDP) proposed for \emph{Intelligent Knowledge Distribution} (IKD) yields per-agent policies for distributing knowledge with other agents based upon the \emph{value of the information} subject to interaction constraints, with a case study in disaster recovery with a heterogeneous multi-robot team.
An optimal POMDP policy is solved through policy iteration and then a combination of Markov chain Monte Carlo (MCMC) and discrete optimization are applied to alter the controller to improve the probability it will not violate resources thresholds over a certain horizon.
Soft constraints are relevant in communication decisions compared to hard constraints for multiple reasons.
Infinite-horizon POMDPs are often represented as a finite state controller \cite{sigaud2010mdp} and evaluating hard constraints on cyclic graphs is intractable due to the stochastic nature of the environment \cite{sigaud2010mdp, hu2007markov}.
In addition, resource availability is often not a hard constraint over short periods of time, as a system may not utilize maximum instantaneous availability or temporary resource over-commitment is permissible (e.g., using queuing protocols such as ZeroMQ).
In \cite{fowler2018icra}, preliminary work in the application of CA-POMDPs to solving IKD was presented and this work expands on it by:
\begin{enumerate*}
  \item Performing a Monte Carlo simulation comparison between intelligent, naive, and greedy communication strategies;
  \item Utilizing Kalman filters as a state estimator to drive information relevance and observation models for IKD;
  \item A discrete optimization algorithm that improves the edge observation transitions of the CA-POMDP; and,
  \item Adapting the CA-POMDP finite state controller policy with Bayesian predictive estimation in response to realized observation probabilities and resource utilization.
\end{enumerate*}

To evaluate the effectiveness of our solution, we require a model that provides a means to gauge relevance of information for communication decisions in our disaster recovery case study.
Kalman filters provide a well established method for state estimation of a ground asset and a means to determine the usefulness of information and the accuracy of our disaster response use case for CA-POMDPs and IKD.
As a result of IKD, the drones monitoring the disaster site were able to satisfy mission objectives, which involved having the ground asset avoid environmental hazards while searching for survivors, while also drastically reducing the consumption of available resources.
In comparison to naive and greedy communication models, the IKD communication model was able to maintain accurate state estimations of the ground vehicle while utilizing significantly less power and bandwidth.

The remainder of this paper is organized as follows.
Section~\ref{sec::related} provides a perspective of our concept in relation to the state of the art.
In Section~\ref{sec::background}, review the fundamental models at the core of our architecture and evaluation case study.
Our approach for Constrained-Action POMDPs is formulated in Section~\ref{sec::capomdp}, which is the basis for providing Intelligent Knowledge Distribution (IKD) in Section~\ref{sec::ikdmodel}.
Section~\ref{sec::exp} describes how the CA-POMDP was validated to perform IKD, with the results reported in Section~\ref{sec::results}.
The paper wraps up with future and continuing research in Section \ref{sec::future}.

\section{Related Work}
\label{sec::related}

Capitan's paper \cite{capitan2013} is a key comparison in analyzing the performance of our paper in a multi-agent coordination setting.
However, \cite{capitan2013} assumes that point-to-point communications are instantaneous and cost-free between nodes, while applying a collaboration policy similar to consensus.
In this work, we instead remove the assumption of instantaneous and cost-free communication to yield multi-agent communication that respects resource constraints.
This results in knowledge distribution that is more nuanced than consensus, i.e., data flooding vs. putting data where it needs to be.

Regarding constraints in decision-making, constrained Markov Decision Processes (MDPs) have been used historically to solve two major drawbacks of standard discrete MDPs: Multiple objectives and limited resources \cite{altman1999cmdp,beynier2006iterative,chamie2015cmdp,feyzabadi2015,fyzabadi2014hcmdp,hansen1998finite,isom2008piecewise,undurti2010online}.
Applying constraints to MDPs has been well established in restricting utilization of states to prevent collisions \cite{chamie2015cmdp}, and has been expanded to hierarchical cases to reduce the complexity of the linear programming formulation \cite{feyzabadi2015, fyzabadi2014hcmdp}.
The primary objective of the Constrained MDP (CMDP) is to find a policy that is within a restricted state, cost, and/or reward structure of an unconstrained MDP.
Constrained POMDPs (CPOMDPs), which provide constraints for partially observable environments, have also been explored using mixed integer linear programming (MILP) \cite{isom2008piecewise,undurti2010online}.
Both  \cite{isom2008piecewise} and \cite{undurti2010online} solve the POMDP using value iteration, though \cite{undurti2010online} also describes a point-based value iteration (PBVI) approach to provide an upper bound to a heuristic search.
An online algorithm has also been proposed \cite{undurti2010online} to address drawbacks related to the propagation of risk aversion.
A finite horizon policy is computed off-line with a ``constraint-penalty-to-go'' for every step of dynamic programming value iteration as defined in \cite{isom2008piecewise} for solving the MILP.

The above approaches to CMDPs and CPOMDPs apply constraints to the state-space of the model and projection into the value space, but our formulation requires \emph{action-based} constraints as we are limiting the utilization of resources that an action consumes which cannot be tied to physical constructs, such as states that represent ``no-fly'' or ``stay-away'' zones.
As many scenarios require an indefinite length of operation and no predefined goal states (as in our case study), our CA-POMDP approaches needs to be solved as an infinite-horizon policy.
Infinite horizon POMDPs are solved by policy iteration with a finite state controller representation \cite{hansen1998solving}, and thus in our context will require analyzing how a cyclic graph utilizes resources with respect to  soft constraints.
To the best of the authors' knowledge, it is not possible to represent soft resource constraints on actions of a cyclic controller in the state or value space of state-of-the-art CMDPs and CPOMDPs.

The modeling of multi-agent systems (MASs) that have common objectives can be represented as a Decentralized Markov decision process (Dec-POMDP).
A Dec-POMDP is a construct that allows multiple independent POMDPs running on different agents to act independently while working towards an objective function that is dependent on all the agents' actions \cite{oliehoek2016concise}.
In brief, the agents only have access to their local observations and as the objective function is dependent on the behavior of all the agents, each agent must maintain a \emph{belief} of the other agents policies.
There are many subclasses of the Dec-POMDP that address joint and local observations, communications, model independence, etc \cite{oliehoek2016concise,mit2015dmuu}.
The Dec-POMDP-Comm is the dominant subclass related to this paper where the agents are reasoning on when to communicate due to delays and costs.
The most commonly seen approach is to make the decision a part of Dec-POMDP model based upon joint state and actions \cite{oliehoek2016concise}.
There is also an approach where there is a centralized POMDP plan that is communicated between the agents which triggers communications when their individual Dec-POMDP plan differs from the centralized one.
Unlike Dec-POMDP-Comm, our CA-POMDP formulation does not utilize communications to inform the joint policy of multiple agents driving communications, instead the model is an \emph{agnostic ``plug-and-play''} approach that separates other collaborative tasks of an agent from communication decisions while enforcing soft resource constraint satisfaction.

Driving communications from the \emph{value of information} similar to CA-POMDP is not unique, but has been the focus of research behind reward shaping and belief-dependent rewards \cite{williamson2009reward,araya2010pomdp}.
These techniques use information-theoretic measures between probability distributions, like KL-Distance, as part of the reward function for belief-dependent rewards \cite{araya2010pomdp} and for determining communication \cite{williamson2009reward}.
\cite{williamson2009reward} focuses on restricting communications to when they are needed but does not provide soft constraint satisfaction to the policy controller which is the focus of this work.
The authors in \cite{williamson2009reward} use the Dec-POMDP-Comm as their basis but change the perspective from a cost to a reward for communication to highlight an opportunity to use a resource.
They state efficient policy generation techniques will be adapted to allow for scalability, where they use the information-theoretic concepts from Dec-POMDP-Value-Comm \cite{williamson2008principled} as a measure of belief divergence.
Alternatively, \cite{spaan2015decision} purposely restructures the action space so they can remain in a classic POMDP problem.
In this paper, we approach the IKD problem similarly to \cite{spaan2015decision} where information rewards drive cooperative perception, but instead of restructuring the action space we restructure the observation model to describe belief on the value of information.

\section{Background}
\label{sec::background}

In this section, we review the principles behind our formulation and provide a foundation of concepts, terminology, and lexicon used in the paper.
Afterwards, we explain how these concepts are integrated into a single framework to provide a Constrained-Action POMDP (CA-POMDP) leading to an \emph{Intelligent Knowledge Distribution} model.

\subsection{Partially Observable Markov Decision Processes}
Markov decision processes (MDPs) are a decision-theoretic approach to determining the best actions over a time horizon based upon the current state $s$ of a system, the probability of transition $T(s'|s,a)$ to another state $s'$ based on action $a$, and the reward $R(s,a)$ associated taking actions in a given state  \cite{sigaud2010mdp, hu2007markov}.
This model is used to determine actions that maximize rewards, i.e., a policy $\pi$, when the rewards are delayed or realized at some future point in time.
If the state of the agent is not fully observable then the model becomes a Partially Observable MDP (POMDP) where belief states, $b \in \mb{B}$, are used to represent the probability distribution of being in any particular state based upon observations \cite{mit2015dmuu}.
These belief states are tied to an observation model, $\mb{O}$, that describes the probability of observations based on the underlying state.
Formulating a POMDP in this manner is often referred to as a \emph{belief-state MDP}.
A POMDP can be described formally by the vector
$ < \mb{S}, \mb{O}, \mb{B}, \mb{A}, \mb{T}, \mb{R}, \mb{\Pi} > $,
where $\mb{S}$ is the set of states an agent can be in, $\mb{A}$ is the set of actions the agent can take, $\mb{T}$ is the set of transition probabilities $T(s'|s,a)$ between states based on an action $a \in \mb{A}$, $\mb{R}$ is the set of rewards $R(s,a)$ for taking an action in a state, and $\mb{\Pi}$ is the set of policies, $\pi$, that are feasible consisting of a set of vectors of actions, $\{\pi_0,\pi_1,\dots,\pi_t\}$.
The objective is to select a policy, $\pi$, that maximizes the expected utility (reward) over time: $\pi^* (s) = \arg\max_\pi U^\pi(s)$ given a utility function $U^\pi(s)$.
Since we do not know the state we are currently in but maintain a belief $b$, an immediate one-step policy would select an action that maximizes the expected reward $\max_a \sum_{s} b(s)R(s,a)$.
If we let let $\alpha_a$ represent $R(\cdot,a)$ as a vector, often referred to as the \emph{alpha vector}, and the current belief state as a vector $\mb{b}$, then an immediate one-step policy would become $\max_a \alpha_a^T \mb{b}$.
An alpha vector thus represents a hyperplane in the belief space that is piecewise linear and convex.

\begin{figure}[!t]
  \centering
  \includegraphics[width=0.66\columnwidth]{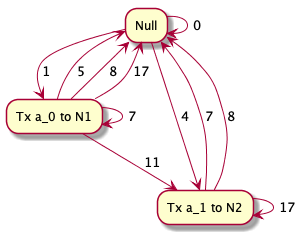}
  \caption{\footnotesize A graphical representation of a Finite-state Controller for a POMDP policy where the vertices $v \in \mc{V}$ represent actions and the edges $e \in \mc{E}$ are transitions based upon environmental observation $o \in \mb{O}$. The example has three actions: (1) transmit message $a_0$ to Node $1$, transmit message $a_1$ to Node $2$, or no transmission $Null$.}
  \label{fig::background::policy}
\end{figure}

In order to represent infinite-horizon POMDP policies, we assume the typical policy representation of a finite state controller (FSC).  In Figure~\ref{fig::background::policy} we show a simple policy, $\pi$, represented as an FSC, which is a directed cyclic graph with the vertices $v \in \mc{V}$ representing \emph{machine states} consisting of an action $a \in \mb{A}$ per vertex and edges $e \in \mc{E}$ representing transitions from an action of machine state $a_i$ to an action of machine state $a_j$ determined by the observation $o \in \mb{O}$ seen after executing that action $a_i$.
An action $a_i$ of machine state $i$ can be the same action $a_j$ of another machine state $j$ such that $\exists \,a_i = a_j: i,j \in \mc{V}(\pi)$ though they will consist of unique alpha vectors, $\alpha_i \neq \alpha_j$.

The policy iteration (PI) algorithm of Hansen in \cite{hansen1998solving} is the basis for improving a FSC through transformations which searches within the policy space until epsilon convergence is observed \cite{hansen1998solving}.
In particular, a dynamic programming update is iteratively used to generate a new set of candidate machine states that the policy improvement loop uses to compare the alpha vectors of existing value function $V$ and the new value function $V'$ and modifies the FSC accordingly.
The set of alpha vectors can be run through a linear programming (LP) formulation to prune the set of alpha vectors that are dominated in the belief space by any combination of other alpha vectors, which is explained in \cite{hansen1998solving}.
We have chosen Hansen's algorithm as the starting point for CA-POMDP for its basic implementation to validate the concept and the availability of point-based policy iteration methods \cite{pineau2003point} for Hansen's approach that reduce the computational complexity.

\subsection{Markov Chain Monte Carlo}
To analyze the probabilistic constraint satisfaction of an FSC policy from an \emph{unconstrained} POMDP, the Markov chain Monte Carlo (MCMC) sampling methodology is applied.
A MCMC is a particle-based approximate inference sampling technique that uses a sequence of samples to iteratively approach a desired posterior distribution \cite{koller2009pgm}.
For CA-POMDPs, the MCMC provides a mechanism to continuously sample an FSC policy seeking a posterior that represents its resource utilization.
When samples are initially likely to be from the prior distribution, the sampling of a sequence of samples will successively approach the posterior, whereas likelihood based approaches are unlikely to account for this fact and place undue weighting on earlier samples.
Another gain in utilizing MCMCs in determining the posterior distributions is the versatility in the type of distribution that can be inferred without being bound to Gaussian distributions or Gaussian approximations.

The Metropolis-Hastings algorithm to solving MCMC was selected because of the difficulty of direct sampling from the probability distribution of a cyclic controller and calculating the normalizing factor of the distribution.
The algorithm uses a random walk approach that either accepts or denies a proposal rather than trying to track importance weights.
The acceptance ratio reduces simply to
\begin{align} \label{eq:mcmc:acceptance}
  \mc{A} = \frac{ P(x|\mu)P(\mu) }{ P(x|\mu_0)P(\mu_0) }
\end{align}
of the proposed posterior distribution, $P(x|\mu)P(\mu)$, over the current posterior distribution, $P(x|\mu_0)P(\mu_0)$.
The most difficult part of calculating the posterior with the Bayes formula, the evidence $P(x)$, is common to both the proposed and current posterior and therefore conveniently cancels out.

If a random number from $[0,1]$ is lower than the acceptance ratio \eqref{eq:mcmc:acceptance}, then the proposed posterior is accepted.
In cases where the proposed distribution is larger than the current distribution, then the acceptance ratio will be greater than one $\mc{A} \geq 1$ and therefore the proposed distribution will always be accepted.
On the contrary when the acceptance ratio is less than one $\mc{A}< 1$, then there is a uniform probability that it will accept the proposed distribution.

\subsection{Graph Theory}
Finally, we require a model for interactions in a decentralized multi-robot system.
Consider a multi-robot system composed of $n$ robots with indices $\mc{I} = \{1,\ldots,n\}$, operating in $\mbb R^d$, each having position $\mb{x}_i \in \mbb{R}^d$.
As an example in determining collaboration between agents, it can be assumed that the robots can intercommunicate in a proximity-limited way, inducing interactions (topology) of a time varying nature.
Specifically, letting $d_{ij} \triangleq \norm{x_{ij}} \triangleq \norm{x_i-x_j}$ denote the distance between robots $i$ and $j$, and $(i,j)$ a link between connected robots, the neighborhood $\mc{N}_i$ of each robot $i$ is defined by the interaction radius $\rho(x_i)$, which has a dependence on the location of robot $i$.
Note that such an interaction radius encodes typical sensing and communication constraints that vary spatially.
The assumed spatial interaction model is formalized by the \emph{directed graph}, $\mc G = (\mc{I}, \mc{E})$ with nodes indexed by the robots, and edges $\mc E \subseteq \mc I \times \mc I$  such that $(i,j) \in \mc E \iff  \| x_{ij} \| \leq \rho(x_i)$ \cite{mesbahi2010graph}.
The interaction graph $\mc{G}$ of a decentralized multi-robot system describes the neighbors $\mc{N}_i$ that an agent $i$ transmits information to in IKD-based collaboration.

\section{Constrained-Action POMDP} \label{sec::capomdp}
The \emph{Constrained-Action POMDP} is a formulation that seeks to find a near optimal policy in a partially observable environment with action-based constraints that are probabilistically guaranteed to stay within specified \emph{soft} limits.
The framework first solves for an unconstrained optimal policy then improves the probabilistic constraint satisfaction of the controller.
Any action performed by an autonomous system will have a distribution representing the utilization of resources, and thus we aim to determine the probability that a series of actions will respect a soft constraint through an analysis of the cumulative distribution function (CDF) of policy resource utilization.
In the formulation of the CA-POMDP, the analysis of probabilistic constraint satisfaction is performed by sampling a policy with Markov chain Monte Carlo (Section~\ref{sec::capomdp::eval}) and improving the constrained policy through discrete optimization (Section~\ref{sec::capomdp::improve}).

\subsection{CA-POMDP Model}
Our goal is to adapt an infinite-horizon FSC of a POMDP policy $\pi$ from policy iteration \cite{hansen1998solving} to probabilistically stay within soft constraints over a desired period of time $T$, i.e., $p(\sum_{t=0}^T u_h(a_t) \leq c_h) \geq \eta_h$, for all constraints $h \in \mathcal{H}$, with resource limits $c_h \in \mb{C}$, where $u_h \in \mb{U}$ is resource utilization used by the action $a_t$.
Policies $\pi$ are represented as a finite-state controller (FSC), since an FSC is a cyclic graph that represents an infinite-horizon POMDP well \cite{bernstein2005bounded}.
The FSC representation can also accelerate the convergence to a policy, since an infinite horizon policy in value iteration is not guaranteed to converge \cite{sigaud2010mdp}.
The CA-POMDP can be described formally by the tuple
\begin{equation}
< \mb{S}, \mb{O}, \mb{B}, \mb{A}, \mb{T}, \mb{R}, \mb{\Pi}, \mc{H}, \mb{C}, \mb{U}, \mb{E}>
\end{equation}
where $\mb{S}$, $\mb{O}$, $\mb{B}$, $\mb{A}$, $\mb{T}$, $\mb{R}$, and $\mb{\Pi}$ are the same as defined in a POMDP,
$\mc{H}$ is the set of resources being used by the agents,
$\mb{C}$ is the matrix defining the constraints for each resource,
$\mb{U}$ defines the utilization of resources for an action $a \in \mb{A}$ and its variation or uncertainty, and
$\mb{E}$ is the matrix of edge observation probabilities for transitioning from a machine state $i$ to another $j$ in an FSC based upon action observation histories (AOH).

\begin{figure}[!t]
  \footnotesize
  \begin{algorithmic}[1]
    \Procedure{ConstraintImp}{$\mb{C}, \mb{U}, \mb{E}, \pi_0$}
      \State $\pi^*, V^*, \Omega \gets$ \Call{POMDP Policy Iteration}{$\pi_0$}
      \State $N_0 \gets$ \Call{ConstraintEval}{$\pi^*,\mb{C},\mb{U},\mb{E}$}
      \If{$N_0 \geq \eta$}
        \State \Return $\pi^*, N_0$
      \EndIf
      \State $\hat{\pi}_0 \gets \pi^*; \quad t \gets 0$
      \Repeat
        \State $t \gets t + 1$
        \State $\Pi' \gets $ \Call{AllFeasibleRegions}{$\hat{\pi}_{t-1}, \Omega$}
        \State $\pi_{\min} \gets \min_\pi \pi \in \Pi'$
        \State $g(\Pi') \gets \left\lVert\right. V^*-$ \Call{PolicyEval}{$\pi_{\min}$} $\left.\right\rVert$
        \State $Q \gets \Pi'$
        \Repeat
          \State $\Pi' \gets \arg\min_{q \in Q} g(q)$
          \State $\pi_{\max} \gets \max_\pi \pi \in \Pi'$
          \State $f(\Pi') \gets \left\lVert\right. V^*-$ \Call{PolicyEval}{$\pi_{\max}$} $\left.\right\rVert$
          \State $N \gets $ \Call{ConstraintEval}{$\pi_{\max}, \mb{C},\mb{U},\mb{E}$}
          \State $\mb{R} \gets $ \Call{BranchController}{$\pi',\mb{U}$}
          \For{$R \in \mb{R}$}
            \State $\pi_{\min} \gets \min_\pi \pi \in R$
            \State $g(R) \gets \left\lVert\right. V^*-$ \Call{PolicyEval}{$\pi_{\min}$} $\left.\right\rVert$
          \EndFor
          \State $Q \gets Q \cup \mb{R}$
          \State $Q.Pop(\Pi')$
          \State \Call{Prune}{$Q$}
        \Until{$Q=\emptyset$}
        \State $\hat{\pi}_i \gets \arg\max_{\pi \in Q} g(\pi)$
      \Until{$\hat{\pi}_i=\hat{\pi}_{i-1}$}
      \State \Return $\hat{\pi}_i, N(\hat{\pi}_i)$
    \EndProcedure
  \end{algorithmic}
\caption{\footnotesize Algorithm for performing constraint improvement of an optimal finite state controller.}
\label{fig::capomdp::code::improve}
\end{figure}

The constraint model consists of a matrix of soft constraints $\mb{C}$ (e.g., power and bandwidth), the utilization of resources $\mb{U}$ per time epoch $\Delta t$ (a decision is made each epoch of time), and the matrix of observation probabilities $\mb{E}$.
Specifically, $\mb{C} \in \Re^{|\mc{H}| \times 2}$ is a matrix representing the desired soft constraints within a specified time period $T$ where each row $h \in \mc{H}$ of $\mb{C}$ is the vector $[ c_h, \eta_h ]$, with $c_h$ the desired upper limit for resource $h$, and $0 < \eta_h < 1$ is the probabilistic constraint satisfaction for that resource based upon operational scenarios.
The resource utilization $\mb{U}$ defines for each resource $h \in \mc{H}$ a matrix of Gaussian probability distribution parameters for all actions, i.e., $u_h(a) \gets \mathcal{N}(\mu_h^a, \sigma_h^a)$.
The matrix $\mb{E} \in \Re^{|\pi| \times |O|}$ consists of edge observation probabilities $p(o|a_{i \in \pi})$ for encountering an observation $o \in \mb{O}$ while at a machine state $i$ of the finite state controller policy $\pi \in \Pi$.
Edge observation probabilities are tied to the FSC policy representation as a means to track the stochastic nature of the environment causing observation transitions from one machine state to another.
Their probabilistic distribution is initially assumed to be uniform and the true distribution is learned online, see Section~\ref{sec::capomdp::learn}.

Our core mechanism in CA-POMDP is the FSC representing the optimal policy calculated using Hansen's POMDP PI algorithm \cite{hansen1998solving}, along with discrete optimization which acts to constrain the policy.
During the discrete optimization phase, CA-POMDP will introduce \emph{constraint states} $j$ into an unconstrained controller to bring the controller within a probabilistic constraint satisfaction $\eta$ (analyzed via MCMC) while minimizing the impact on the optimal unconstrained value function $V^*$.
To effectively alter the controller with constraint states, the non-dominant alpha vectors per action, defined as set $\Omega$, are maintained during the dynamic programming updates for unconstrained policy optimization.
These alpha vectors are not dominant at any dynamic programming update, but may utilize fewer resources and therefore should be considered for altering the controller for probabilistic constraint satisfaction.

\subsection{Constraint Improvement} \label{sec::capomdp::improve}

Improving the constraint satisfaction of an $\epsilon$-optimal unconstrained policy is solved using a branch and bound (BnB) algorithm variant of \cite{clausen1999branch}, a well-known technique for discrete and combinatorial optimization.
The underlying goal of the BnB is to determine a set of constraint states, our previously stored set of alpha vectors $\Omega$, to ``inject`` into the optimal controller which utilize fewer resources than the existing machine states in the FSC while trying to minimize the loss in the expected value from $\epsilon$-optimal.

\begin{remark}
  By calculating the optimal policy first, we also have an initial optimal policy to recalculate from as resource utilization and observation edge probabilities are learned online (Section \ref{sec::capomdp::learn}). 
\end{remark}

\begin{figure}[!t]
  \includegraphics[width=\columnwidth]{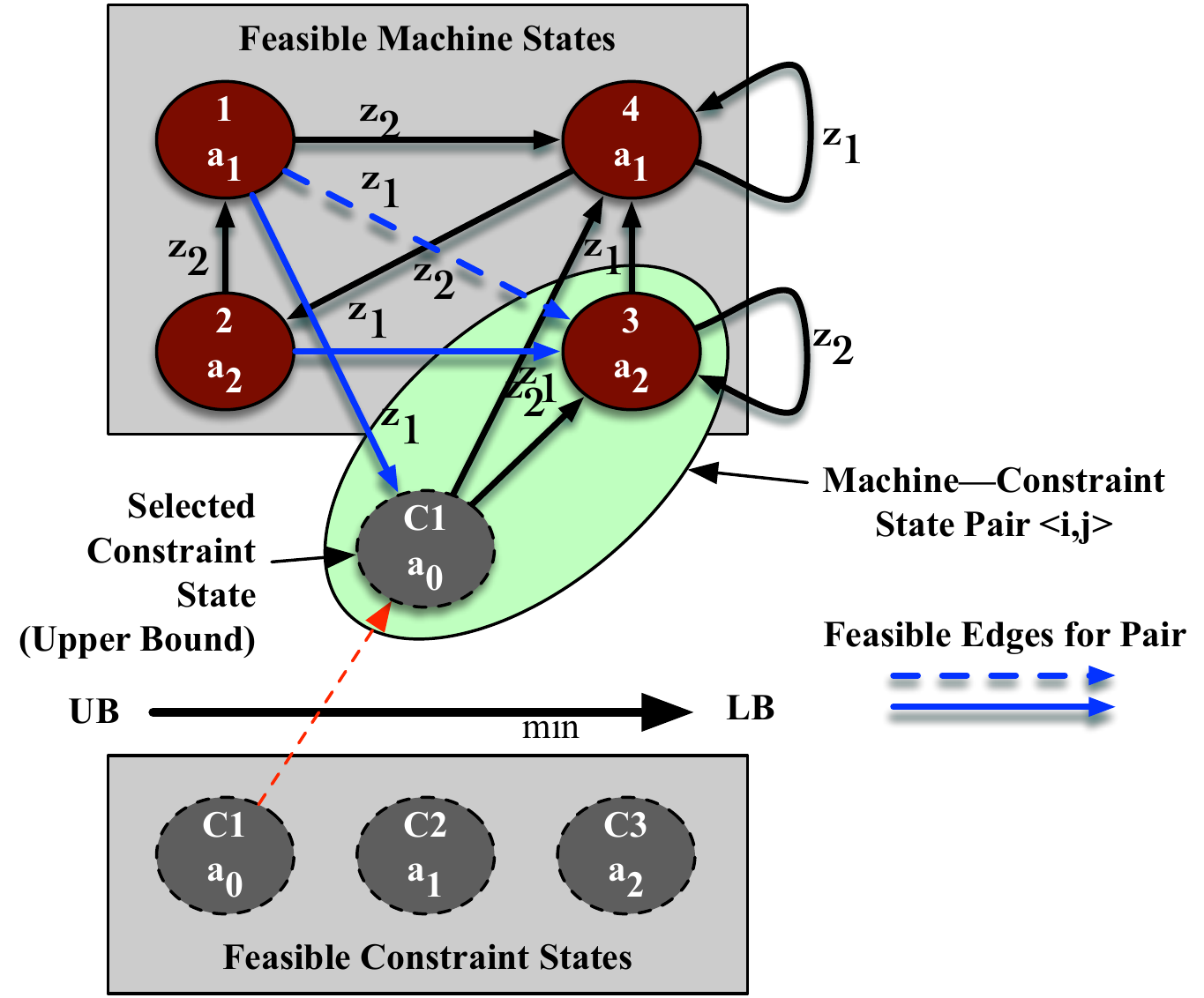}
  \caption{\footnotesize An example of the feasible regions for a finite state controller for use in the Branch and Bound Discrete Optimization Algorithm.}
  \label{fig::capomdp::feasibility}
\end{figure}

The algorithm (Figure~\ref{fig::capomdp::code::improve}) initializes with the $\epsilon$-optimal FSC policy $\pi^*$ calculated from policy iteration which is immediately checked for probabilistic constraint satisfaction (Line 3 of Figure~\ref{fig::capomdp::code::improve}).
If the policy is non-compliant, the root of a BnB Tree $Q$ (Line 10) is initialized with the \emph{feasibility space} of modifications that can be made to the unconstrained policy $\pi^{*}$, i.e.,  an ordered combinatorial set of all of machine states, constraint states, and edge redirection variables, see Figure~\ref{fig::capomdp::feasibility}.

The first variable in the feasibility region is the set of existing machine states $i$ in the policy $\hat{\pi}_{t-1}$, shown in the upper gray box of Figure~\ref{fig::capomdp::feasibility}, which is initially $\pi^*$.
Each machine state's dominant region $V_i b_i$, where $b_i$ is the region in the belief space $b$ that machine state $i$ has dominance over all other machine states, is used in creating an ordered set of machine -- constraint state pairs, marked in  Figure~\ref{fig::capomdp::feasibility}.

The set of alpha vectors $\Omega$ saved from the policy iteration algorithm define the second set of variables in the feasibility region.
Each alpha vector becomes a potential constraint state $j$, shown in the lower gray box of Figure~\ref{fig::capomdp::feasibility}, that can be introduced into the controller and will result in a loss of value depending on the value of the new alpha vector and chosen edge redirection (the third feasibility region variable).
A feasible constraint state $j$ is an alpha vector whose action $a_j$ utilizes less resources $U(a_j) < U(a_i)$ than machine state $i$.
The upper $f(\pi)$ and lower bound $g(\pi)$ of a constraint state and machine state pair $<i,j>$ provides the BnB maximum (UB) and minimum (LB) value difference between the constraint state's alpha vector $V_j b_i$ and the machine state's $V_i b_i$ for the machine state's dominant region $b_i$: $\norm{V_i b_i - V_j b_i}$.

The third and final variable in the feasibility region is the set of viable edge redirection probabilities.
During the branch and bound, the feasible constraint states for a particular machine state pair $<i,j>$ are injected into the controller and the edges leading to this new constraint state from other machine states already in the controller are selected as either maintaining their transition to the existing machine state $i$ or redirecting to the new constraint state $j$.
The edge redirections are solved for last and until then a simple assumption is made for the calculation of upper and lower bounds for machine--constraint state pairs.
The lower bound function assumes that none of the edges are redirected which results in a controller that has no value impact.
The upper bound assumes all  edges are redirected to the constraint state, which causes the greatest impact on the value function as the original machine state that dominated has been completely substituted or replaced by the constraint state that maximizes $\norm{V_i b_i - V_j b_i}$.

The blue lines in Figure~\ref{fig::capomdp::feasibility} indicate feasible edges that could be redirected or maintained for the $<3,C1>$ machine -- constraint pair.
The constraint state inherits the outgoing edges of the original machine state $i$ since the outgoing edges already transition to dominant machine states after a belief update, and any constraint state introduced into the controller has already been determined to be non-dominant.
Optimally redirecting an edge from another machine state to the new constraint state is not trivial due to the combinatorial explosion of redirection options, which drastically reduces computational efficiency.
Instead, we allow the BnB to select from a finite set of the probabilities of redirecting an edge from an original machine state to the new constraint state, i.e., a vector of ordered probabilities $\overrightarrow{P} = [p_0, p_1, \dots , p_n]$ where $p_0 < p_1 < \dots < p_n$ and $0 < p_i < 1$.
The BnB algorithm searches for the correct probability $p_i \in \overrightarrow{P}$ of edge redirection, yielding  a solution that satisfies probabilistic constraints while minimizing the impact to the value of the constrained controller.
As the edge redirection probability increases, more edges from an existing machine states $k$ in the controller to the machine state $i$ will be redirected to the constraint state $j$.

Formally, the objective of the Branch and Bound algorithm of Figure~\ref{fig::capomdp::code::improve}, is to minimize the impact on the constrained controller value $V(\hat{\pi})$ compared to the optimal controller $V^*$ while ensuring that the probabilistic constraint satisfaction $N(\hat{\pi})$ is greater than or equal to the soft constraints $\eta \in \mb{C}$.  That is,
\begin{align}
    \begin{array}{r l}
      Objective: & \min_{\hat{\pi}} \norm{V^* - V(\hat{\pi})} \\
      s.t.: & N(\hat{\pi}) \geq \eta \quad \eta \in \mb{C}.
    \end{array}
\end{align}
During each step of the algorithm, a set of feasible solutions $\Pi'$ is removed from the tree $Q$ based on a priority metric (best-first search) provided by the lower bound function $g(\Pi')$ (Line 12).
The branching function (line 18) simply divides the region of feasible solutions in half, similar to constrained integer programming \cite{taylor2009integer}, creating two new nodes $\Pi'$ in the tree $Q$ until there is no region to divide creating a ``leaf'' node.
The probabilistic constraint satisfaction $N$ of the upper bound policy $\pi_{\max}$ of a node $\Pi' \in Q$ is evaluated via MCMC (Line 17), described in the next section, since we are interested in the best satisfaction the worst controller can provide.
The pruning function (Line 19) removes any node in the tree where:
\begin{enumerate*}
  \item the lower bound of that node is greater than the upper bound of any other $g(\Pi_i') < f(\Pi_j') \quad \forall (i,j) \in Q_{i \neq j}$; or,
  \item the upper bound does not meet the constraint satisfaction $N_i < \eta \quad \forall i \in Q, \eta \in \mb{C}$.
\end{enumerate*}

There are two loops in the constraint improvement algorithm with the inner loop terminating when there are no further nodes to consider in the tree $Q$ (Line 20).
The outer loop (Line 22) introduces a single constraint state during each iteration until the desired probabilistic constraint satisfaction has been achieved.
Afterwards, a constraint state can be introduced into the finite state machine with an objective to increase the value function without violating the constraint.
Once there are no changes to the controller to improve constraint satisfaction or value function, the function returns the new controller.

\begin{figure}[!t]
  \includegraphics[width=\columnwidth]{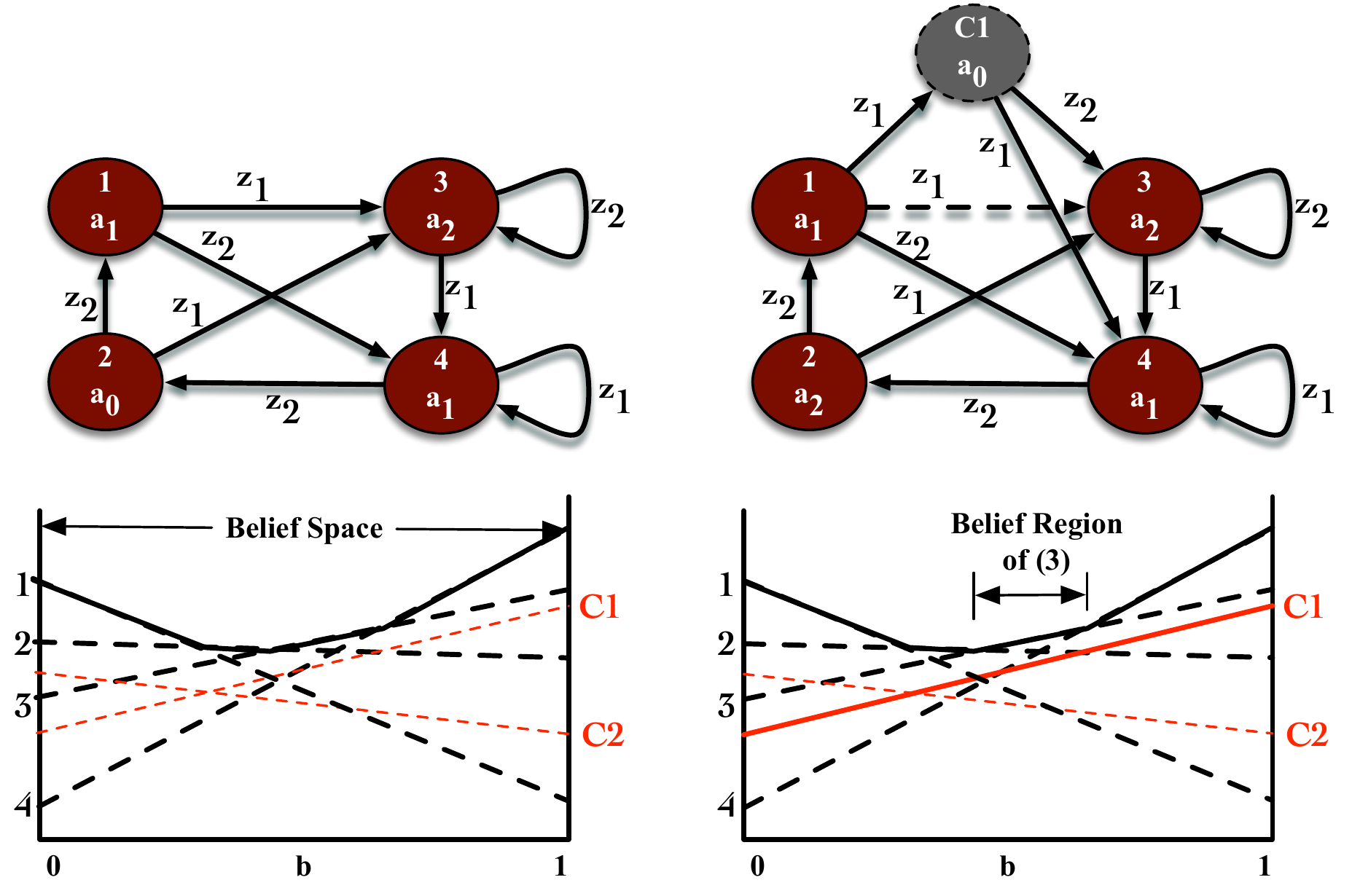}
  \caption{\footnotesize The constraint $C1$ has been injected into the controller to reduce the resource utilization of machine state $3$. The edge redirection function has selected the $1 \rightarrow 3$ to be reconnected as $1 \rightarrow C1$ and $C1$ inherits the outgoing edges of $3$. Notice the loop for observation $z_2$ at $3$ becomes a transition of $C1 \rightarrow 3$}
  \label{fig:ci:example}
\end{figure}

Figure~\ref{fig:ci:example} provides an example of a constraint improvement step with the original optimal FSC shown on the left with machine states $1$ through $4$.
Notice that the machine states $1$ and $4$ have the same action, $a_1$, which is an indicator that the same action during policy iteration was dominant in two different belief regions.
During the constraint improvement, the branch and bound is analyzing the possibility of a constraint state $C1$ concurrent with machine state $3$ utilizing action $a_0$, which utilizes fewer resources than action $a_2$.
$C1$ was initially selected because $C1$ has the greatest value in the same belief region as $3$, since in a best-first search it minimizes $\norm{V_3 b_3 - V_{C1} b_3}$ and can be seen in the red line of the right graph of Figure~\ref{fig:ci:example}.
If $C1$ fails to satisfy the constraints, then the algorithm will examine using $C2$ assuming $C2$ utilizes fewer resources then $C1$.

The edge redirection function has selected to redirect the edge $(1,3)$ to constraint state $C1$ and $C1$ inherits the outgoing edges from machine state $3$.
It is important to note that the loop at machine state $3$ for observation $z_2$ in the original controller does not create a loop in $C1$ for $z_2$ but is redirected to the original machine state since $3$.  This occurs since machine state $3$ is the dominant machine state for the belief update of $z_2$, thus the BnB procedure avoids a non-optimal loop in the constrained controller.

\subsection{Constraint Satisfaction Evaluation} \label{sec::capomdp::eval}

Evaluating constraint satisfaction, shown in Figure~\ref{fig::capomdp::code::eval}, is accomplished by random sampling of the finite state controller to estimate the resource utilization through a Markov chain Monte Carlo (MCMC) Metropolis-Hastings algorithm.
Using the length of time or epochs over which the constraints are being applied $T$, the finite state controller is sampled $T/\Delta t$ times, where an action is taken every $\Delta t$.  Starting from a random initial machine state, this sampling process follows the FSC edge transitions around the controller and records the resource utilization per action by sampling from distributions $\mathcal{N}(\mu_h^a, \sigma_h^a)$, for each resource $h \in \mc{H}$.
The samples are combined into an aggregate sample $x$ that represents the total mean and deviation of resource utilization for the entire time period $T$.
This is repeated from the same initial machine state to gather a sequence of samples $\mb{x}(\alpha)$ until the chain has approached a representative posterior $P^{T}(\alpha)$ of the FSC resource utilization for that particular initial machine state, where $\alpha$ represents the distribution of sample data.
The posterior has been considered ``mixed'' when the variational distance is within epsilon \cite{koller2009pgm} (Line 13)

\begin{align} \label{eq::capomdp::vardist}
  \mathbf{D}(P^{(T)} ; \mb{x}(\alpha) ) = \max_{\alpha} |P^{(T)}(\alpha) - \mb{x}(\alpha)| \leq \epsilon_D.
\end{align}

\begin{figure}[!t]
  \footnotesize
  \begin{algorithmic}[1]
    \Procedure{ConstraintEval}{$\pi^*,\mb{U}, \mb{A},\mb{C},\mb{B}, l$}
      \State $\mb{P}^{(T)} \gets \emptyset$
      \Repeat
      \State $b_0 \gets$ \Call{Random}{$b \in \mb{B}$}
      \State $i^* \gets \arg\max_{i \in \pi^*} V_i \cdot b_0$
      \State $\mb{x} \gets \emptyset$
      \While{True}
        \For{$t \gets 1$ to $T$ step $\Delta t$}
          \State $\mb{x} \gets \mb{x} \cup U(a_{i^{*}} \in \mb{A})$
          \State $i^* \gets$ \Call{RandomEdge}{$p(o|i^*), P^{(T)}(\alpha)$}
        \EndFor
        \State $\mb{x}(\alpha) \gets$ \Call{GaussianFit}{ $\mb{x}$ }
        \If{$D(P^{(T)};\mb{x}(\alpha)) \leq \epsilon_D$} \Comment{Equation \eqref{eq::capomdp::vardist}}
            \State \textbf{break}
        \EndIf
        \If{ \Call{Random}{$0,1$} $\leq \mc{A}$ } \Comment{Equation \eqref{eq:mcmc:acceptance}}
          \State $P^{(T)}(\alpha) \gets \mb{x}(\alpha)$
        \EndIf
      \EndWhile
      \State $\mb{P}^{(T)} \gets \mb{P}^{(T)} \cup P^{(T)}(\alpha)$
      \Until{$Cov \left[ \mb{P}^{(T)}[m]; \mb{P}^{(T)}[m + l] \right] < \epsilon_{cov}$} \Comment{Equation \eqref{eq::capomdp::cov}}
      \State \Return $\int_{-\infty}^{\eta \in \mb{C}} \mb{P}^{(T)} dx$
    \EndProcedure
  \end{algorithmic}
\caption{\footnotesize Algorithm for Evaluating a Finite State Controller for its Probabilistic Constraint Satisfaction.}
\label{fig::capomdp::code::eval}
\end{figure}

To determine when the algorithm needs to stop taking samples from random initial machine states, the algorithm continues until the autocovariance of the lagged MCMC distributions, a generalization of the central limit theorem to Markov chains, has converged \cite{koller2009pgm} to within $\epsilon_{cov}$

\begin{align} \label{eq::capomdp::cov}
  &Cov \left[ \mb{P}^{(T)}[m]; \mb{P}^{(T)}[m + l] \right] \approx \\
  &\frac{1}{M-l} \sum_{m=1}^{M-l} (\mb{P}^{(T)}[m] - \hat{\mb{E}}(m)) (\mb{P}^{(T)}[m+l] - \hat{\bf{E}}(m + l)), \nonumber
\end{align}

where $\mathbf{M}$ is the set of samples collected, $l$ the number of samples to lag, $\mb{P}^{(T)}$ is the vector of $P^{(T)}(\alpha)$, and $\hat{\mathbf{E}}(\cdot)$ is the unbiased estimator $\frac{1}{M} \sum_{m=1}^M P^{(T)}(m)$.
%

Once the MCMC has terminated, the resource utilization posterior distribution is used to determine the probability that the current FSC satisfies the soft resource constraints $\eta$.

\subsection{Learning Edge Probabilities and Resource Utilization} \label{sec::capomdp::learn}
When sampling the controller during the evaluation of constraint satisfaction, an observation edge $o$ is randomly chosen during each of the samples with a probability $p(o|a_i)$ depending on the action $a_i$ of machine state $i$.
This Monte Carlo sampling of observation edges through the controller is initially a uniform \emph{a priori} distribution, but the actual conditional probability of an observation given an action during operation needs to be learned so that the controller can be updated to ensure proper constraint satisfaction as the environment changes.
Observations $o$ given an action $a$ are described by a categorical distribution (aka, a generalized Bernoulli or multinoulli distribution), which is used to represent the likelihood of some finite set of possible observations.
The conjugate prior of a multinoulli distribution is a Direchlet distribution, which is also a Jeffrey's Prior for an N-sided die with biased probabilities, $\overrightarrow{\gamma} = (\gamma_1, \dots, \gamma_N)$.
The closed form solution for calculating the posterior distribution for the categorical distribution and a Direchlet prior is
\begin{align}
  p(o|A) &= \frac{c(o,a) + \alpha(o,a)}{\sum_{j=1}^{|c(j)|} c(o,j) + \sum_{j=1}^K \alpha(o,j)} ,
\end{align}
where $c$ is the \emph{a priori} count for an observation $o$ with state action $a$, $\alpha$ the observed occurrence of observation $o$, and $K$ the total number of occurrences seen.
This method allows a deployed system to track the relationship between actions and observations over time and then recalculate a constrained policy to maintain constraint satisfaction or maximize the value of the controller if it is utilizing fewer resources than expected.

Another consideration in adapting the controller for constraint satisfaction is validating the \emph{a priori} resource utilization models are still applicable during operation.
Situations in the field change over time and the \emph{a priori} models used for resources will not be valid during the entire operational life of the autonomous system.
As an example, as the battery of an autonomous system is drained during its operation, an action may consume more battery power than when the battery is fully charged.
To track the resource utilization, a simple Bayes estimator with Gaussian priors and likelihood is used to track the current resource utilization per action.

It is unrealistic to continuously update the controller with every edge observation or resource distribution change due to limitations in computational resources and to prevent short-term instabilities or unpredictability in the controller.
However, we still need to adapt online to ensure constraint satisfaction or improve the FSC value.
To define an appropriate trigger for recomputing a new controller, the information-theoretic concept of Variation of Information is utilized.
When the variation of information
\begin{align}
  d(X,Y) &= H(X|Y) + H(Y|X) \\
  H(X|Y) &= - \sum_{i,j} p(x_i,y_i) \log \frac{ p(x_i,y_i) }{ p(y_i) } \; \forall i \in X, j \in Y \label{eq:varinfo}
\end{align}
exceeds a desired threshold, the algorithm will recompute a constrained controller from the precomputed optimal controller, where $X$ is either the \emph{a priori} probability distribution of edges $\mb{E}$ or resource utilization $\mb{U}$ and $Y$ is the associated probability distribution of the learned distribution.
When the learned distributions are significantly different than those previously used to compute a constrained FSC, we recompute the controller with the new learned probability distributions.

\section{Intelligent Knowledge Distribution Model} \label{sec::ikdmodel}
Intelligent Knowledge Distribution (IKD), in this paper, is applying a Constrained-Action POMDP to control communications between multiple independent agents that must stay within quality of service limitations.
The goal of IKD is to answer the questions:
\begin{enumerate*}[label=(\roman*)]
  \item What information should we send,
  \item when should we send it, and
  \item to whom should we send it to?
\end{enumerate*}

\subsection{Action \& State Model}
The IKD Model builds on the CA-POMDP formulation by extending the CA-POMDP tuple to include $<\mb{\Sigma}_i, \mc{I}, \mc{N}_i, \mb{F}>$, where we assume each agent independently runs CA-POMDP.
Each agent $i$ has a set of neighbors $\mc{N}_i$ in the environment that can actively collaborate.
$\mb{\Sigma}_i$ is the alphabet of communications $\sigma_k \in \mb{\Sigma}_i$ that agent $i$ can transmit to a neighbor $j \in \mc{N}_i$.
$\mb{F}$ is a set of states $s \in \mb{S}$ that place mission objectives at risk of failure, which we will use to drive information relevance.
We define the set of actions $\mb{A}_i$ for an agent $i$ that indicate the decision to transmit the $k$th element of information to agent $j \in \mc{N}_i$, denoted as action $a_k^j$ related to the information $\sigma_k$:
\begin{equation} \label{eq:ikd:actions}
  \mathbf{A_i} = \left\{\cup_{ \sigma_k \in \mb{\Sigma}_i, j \in \mc{N}_i,  } \quad a_k^j \rightarrow \{0,1\}  \right\} \; \cup \varnothing
\end{equation}
along with the single $\varnothing$ (or Silence) action to not transmit at all.

The states of the IKD model are defined by levels of relevance $S_r$ of locally observed information for each agent $i$ (e.g., $S_r = \{LOW, MEDIUM, HIGH\}$), and levels of collaboration $S_c$ for each agent $i$ with its neighbors $j \in \mc{N}_i$ (e.g., $S_c = \{LOW, MEDIUM, HIGH\}$).
Relevance is described by a set of discrete states that indicate how important information is to global mission objectives, and is calculated per-agent based on local observations only.
The confidence of an agent in its current collaboration with its neighbors $j \in \mc{N}_i$ is a set of discrete states indicating a level of confidence that the current level of communication will maintain the ability for the agents to achieve global mission objectives by sharing local observations.
Formally, the set of states for an agent $i$ in the IKD model is:
\begin{equation} \label{eq:ikd:states}
  \mathbf{S_i} = \{ s_r^{k=1}, \dots, s_r^{k=|\Sigma_i|},  s_c^{j=1}, \dots,  s_c^{j=|\mc{N}_i|} \}
\end{equation}
with relevance states $s_r^k \in S_r$ for each of information element $k$ and a set of collaboration states $s_c^j \in S_c$ with a neighbor $j \in \mc{N}_i$ indicating whether the neighbor is ``up to date'' with relevant information.
The states in an agent's state set therefore becomes a representation of how important information is and how timely the agent's information is for neighboring agents.

\subsection{Reward Model}
The reward functions for IKD need to be defined specifically by the model designer or learned online;  what follows are simply basic guidelines for reward formulation.
The reward for relevance $\hat{R}_r (s,a)$ is a normalized reward based upon the product of
\begin{enumerate*}
  \item the likelihood $\mc{L}$ that the next state $s'$ is not a critical state $f \in \mb{F}$ where global mission objectives are at risk when taking action $a$ in state $s$, which is either known \emph{a priori} or learned online, and
  \item the information-theoretic metric on the maximal value of the information $\rho(k)$ an agent could convey to a neighbor:
\end{enumerate*}
\begin{align}
  \hat{R}_r (s,a) = \mc{L}(s' \notin \mb{F}|a,s) \max_{\sigma_k \in \Sigma_i} \rho(k).
\end{align}
The basic concept is to increase the reward for communications when the mission objectives are at risk and the value of the local information is high.

\begin{remark}
The utilization of a reward function with the components $\rho$ and $\mc{L}$ by its nomenclature appears similar to belief-dependent rewards \cite{williamson2009reward}, however we argue here why this is not the case.
The reward functions are not tied to a state belief but instead are calculated for each state of the IKD model and assume a linear function between states, as seen with any belief-state MDP \cite{oliehoek2016concise, mit2015dmuu}.
The reward functions can be learned online and the FSC policy recalculated but they will remain piecewise linear and convex within the hyperplane of belief space and do not change as the belief does.
\end{remark}

Collaboration is a normalized reward $\hat{R}_c (a,s)$ based upon the product of the proximity of a quantized state to approaching the heuristic bound where collaboration will diverge and become unbounded $\mc{Q}(\cdot)$, as with intermittent communication in controls with a Kalman filter's estimation error covariance \cite{sinopoli2004kalman}, and the maximal value of information that can be shared with a particular neighbor:
\begin{align} \label{eq:ikd:reward:collab}
  \hat{R}_c (a,s) &= \frac{1}{|\mc{N}_i|} \sum_{j \in \mc{N}_i} \mc{S}( \mc{Q}(\lambda_j | a, s) ) \cdot \max_{\sigma_k \in \Sigma_i} \rho(k) ,
\end{align}
where $\mc{S}( \cdot )$ is the sigmoid function,
$\mc{Q}(\lambda_j|a,s)$ is the probability that the communication action $a$ with neighbor $j \in \mc{N}$ will approach the proximity of an unbounded heuristic $\lambda \in \mb{\Lambda}$,
$\mb{\Lambda}$ is the limits of the upper $\bar{\lambda}$ and lower $\underline{\lambda}$ bound heuristic of a critical probability $\lambda_c$ that is dependent on the model (See Section~\ref{sec:exp:filter}), and
$\max_{\sigma_k \in \Sigma_i} \rho(k)$ is the best value of information $\rho(k)$ that an agent $j \in \mc{N}$ could receive from $i$.
Thus, as the lack of communications drives a global objective towards an unbounded heuristic (e.g., a diverging Kalman filter estimate due to intermittent communications) and the value of information for neighboring agents increases, the reward for communication increases.
The final reward ${R}(s, a)$ is the sum of the collaboration and relevance rewards with one exception.
If the action is not to transmit ($a=\varnothing$), the reward is zero for all states:
\begin{align}
  R(s, a) &= \left\{
    \begin{array}{l l}
      \hat{R}_r + \hat{R}_c & if \: a \neq \varnothing \\
      0 & if \: a = \varnothing
    \end{array}
  \right. .
\end{align}
This at first appears counter intuitive, because the action to not transmit will be dominated by all other actions which is intentional.
If the agent stops transmitting then it always is going to lead to information loss and therefore provides no benefit not to continuously transmit.
It is the constraints to preserve resources that will introduce a do not transmit action or one that utilizes fewer resources into the controller as a viable constraint state to satisfy soft resource constraints.

\subsection{Transition \& Observation Models}
As with the rewards model, transitions and observations are divided into relevance and collaboration components.
Transition probabilities from one state $s$ to another state $s'$ based upon an action $a$ are assumed to be independent to reduce computational complexity and therefore are the product of the independent transitions of the components
\begin{align}
  T(a, s, s') = \prod_{i=1}^n \prod_{\sigma_k \in \Sigma_i} T(a, s_r^k, s_r^{k'}) \cdot \prod_{j \in N_i} T(a, s_c^j, s_c^{j'}) \label{eq:cpomdp:trans}
\end{align}
where $T$ is the transition matrix, and $s_r$ and $s_c$ are the relevance and collaboration state components.
The component transition functions $T(a, s_r, s_r')$ and $T(a, s_c, s_c')$ are determined by the knowledge being shared and are discussed for our use case in Section~\ref{sec::exp::model}.

The observation probabilities are similarly constructed as the transition probabilities with relevance and collaboration components as follows:
\begin{align}
  O(a, s, o) = O(a, s_r, o_r) \cdot \prod_{j \in \mc{N}_i} O(a, s_c^j, o_c^j) \label{eq:cpomdp:obs}
\end{align}
where $O$ the observation matrix, and $o_r$ and $o_c$ are the relevance and collaboration components of observations.
\footnote{In future work, this assumption will also be relaxed as with the transition assumption of independence.}

\begin{remark}
    Though the relevance of the information and the collaboration between agents are conditionally dependent in reality, the formulation of the model as a decentralized POMDP reduces the dependence with the focus on local observations allowing to validate the approach before addressing a conditional dependence transition and observation model.
\end{remark}

As discussed in related work, the observation model is key in restructuring the POMDP to avoid belief-dependent rewards and its associated solution methods.
In the construction of an IKD model, an observation model should be designed to provide the ability to map observations to the appropriate belief space for relevance and collaboration.
Any mapping $\upsilon$ from observations in the environment to an observation state is the mapping of a continuous observation space to discrete observations through  cluster-based techniques, such as K-medoids \cite{murphy2012machine} or DBScan.
See Section~\ref{sec::exp::model} for how an observation model was constructed to address IKD with asset localization and Kalman Filters.

\section{Experimental Model} \label{sec::exp}
The use case for validating the Constrained-Action POMDP for \emph{Intelligent Knowledge Distribution} (IKD) was the aerial monitoring of ground assets during a disaster response to ensure they safely avoid hazards and dangerous situations they may not be aware of from their perspective on the ground.
Though a ground vehicle could potentially have accurate location information from GPS, we consider the asset tracking viable, because:
\begin{enumerate*}
  \item There are situations where GPS information in a disaster site is either unavailable or degraded due to obstructions or other factors;
  \item The actual location of the hazards are not known and the monitoring drones need to maintain accurate relative positioning of the ground asset; and,
  \item The ground asset needs to update its motion planning to optimize for known hazards as they are estimated by monitoring drones.
\end{enumerate*}

\begin{figure}[!t]
  \centering
  \includegraphics[width=0.75\columnwidth]{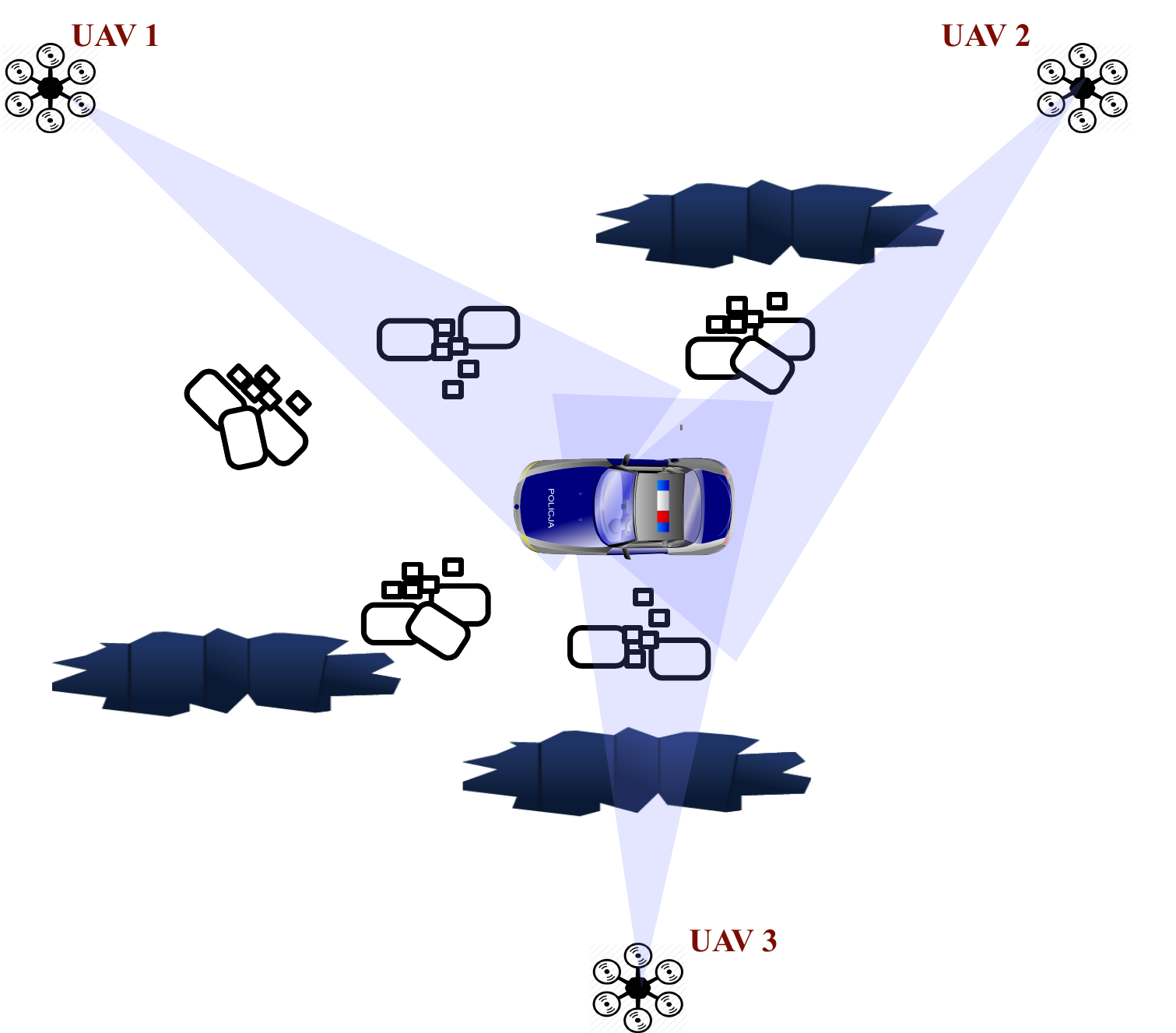}
  \caption{\footnotesize The conceptual diagram of the situation being modeled with an autonomous emergency vehicle traversing a disaster site with blocked views of potential hazards (e.g. sinkholes)}
  \label{fig:scenario}
\end{figure}

In this context, our IKD agents are unmanned aerial vehicles (UAVs) performing a continuous predefined search pattern around the disaster site while communicating across a first responders mobile ad-hoc network (MANET) as shown in Figure~\ref{fig:scenario}.
In this initial study to validate the CA-POMDP and IKD concepts, the agents are only tracking a single ground asset and need to determine what sensor information needs to be shared between them to ensure they maintain situational awareness of the hazard risk.
They also provide a relative position warning to the ground asset so that it can update the edge costs of its motion planning algorithm appropriately, e.g., D$^{\star}$-Lite \cite{koenig2002d}.

The IKD model, as described in Section~\ref{sec::ikdmodel}, is a combinatorial problem between the neighboring nodes, the \emph{heterogeneous} sensor data, the relevance of the information, and the current level of collaboration.
Since an agent can only communicate with a single node at a time due to routing protocols with WiFi Mesh Networking, it needs to determine the risk it believes an asset is under (relevance) and how confident it is in the data it has received so far from other nodes (collaboration) to ascertain the appropriate information to send and to whom to maintain accurate situational awareness, all while not overly consuming the limited resources of bandwidth in the MANET and power available from the UAV battery.

\subsection{Asset Localization \& Kalman Filter} \label{sec:exp:filter}
An unscented Kalman filter (UKF) was utilized in performing localization of the ground vehicle from multiple drones, where each drone is providing a bearing and distance to the ground vehicle with variances per the capabilities of the sensor packages that are aboard that particular drone.
Kalman filters (KF) are a common approach to guidance, navigation, and control of vehicles and robots \cite{musoff2009fundamentals} as a linear quadratic estimator of the state of a system from a series of measurement or observation samples over time.
The Unscented Kalman filter (UKF) uses a deterministic sampling technique known as an unscented transform around a minimal set of sample points about the mean, which performs well in highly non-linear systems as compared to Extended Kalman Filters (EKF) \cite{julier1997new}.
A nonlinear Kalman filter was necessary due to
\begin{enumerate*}
  \item the nonlinearity of the motion model with arcs of motion depending on the angle of the front wheel and
  \item the nonlinearity of triangulating the position of the ground vehicle from drones with sensor bearings and distances.
\end{enumerate*}

The motion model $f(x,u)$ of the ground vehicle is a standard \emph{bicycle model} with static rear wheel and variable front wheel.
The state model $\mb{x}$ maintains information on the position and orientation of the ground vehicle as the vector $\mb{x} = [ x \quad y \quad \theta ]^T$, where $(x,y)$ is the location of the ground vehicle in the disaster site as measured in meters and $\theta$ is the bearing in radians.
The state transition function $\bar{x}$ is a non-linear motion model $f(x,u)$ with Gaussian white noise $\mc{N}(0,Q)$ defined as
\begin{align}
  \bar{x} &= x + f(x,u) + \mc{N}(0,Q),
\end{align}
where $\mb{u}$ is the command input $[v \quad \alpha]^T$ to the ground vehicle control system and is defined by the linear velocity $v$ and the steering angle $\alpha$ of the ground vehicle.
The measurement model $\mb{z}$
\begin{align}
  \mb{z} &= h(\mb{x},\mb{P}) + \mc{N}(0,R) \\
  h(\mb{x},\mb{P}) &= \left[ \begin{array}{c}
    \sqrt{(p_x-x)^2 + (p_y-y)^2} \\
    \tan^{-1} \left( \frac{y - p_y}{x - p_x} \right) - \theta
  \end{array} \right]
\end{align}
involves the bearing and distance to the ground vehicle $\mb{x} =[x,y]$ from the current observation location of drone $[p_x,p_y]$, in which the bearing and range measurement noise
\begin{align}
  \mb{R} &= \left[
  \begin{array}{c c}
    \sigma_{range}^2  & 0 \\
    0 & \sigma_{bearing}^2
  \end{array} \right]
\end{align}
is assumed to be independent and may or may not have line of sight to the target.
In cases where a drone does not have a line of sight of the ground asset nor any observation communications from another drone, this is a missing measurement.

For the UKF, the unscented transform uses a particle-based technique, which requires the ability to calculate the mean of the particles.
Calculating the mean of the positions is a simple mathematical average, whereas calculating the mean of the bearing is
\begin{align}
  \bar{\theta} &= \atantwo \left( \frac{\sum_{i=1}^n \sin \theta_i}{n}, \frac{\sum_{i=1}^n \cos \theta_i}{n}  \right).
\end{align}

There are two occurrences that need to be addressed in the application of Kalman Filters because of intermittent communications and limited line of sight: missing measurements and missing observations.
In missing measurements, there is no information at a discrete time step to update the state estimate and the predicted measurement is propagated forward as an observation \cite{sinopoli2004kalman,thomas2018fusion}.
For missing observations, there are observations at a given time step but not of all the state information.
In the case of geolocation through triangulation, a bearing and distance measurement is not available to accurately pinpoint the robot, but there are enough observations to get a less accurate measurement.
With UKF, the unscented transform is performed as usual, because the model can be constructed ``on-the-fly'' to replicate the variance that would be seen with the lack of observation for a fully accurate triangulation by adjusting the measurement noise, $R$ \cite{nastasi2018autonomous}.

If the probability of arrival $\lambda$ of an observation in a Bernoulli process is less than a critical probability $\lambda \leq \lambda_c$ for a Kalman Filter, then the expectation of the estimation error covariance is unbounded \cite{sinopoli2004kalman}.
The lower bound $\underline{\lambda}$ of the critical probability $\lambda_c$ has a closed form solution but the upper bound $\bar{\lambda}$ requires solving a linear matrix inequality (LMI) \cite{sinopoli2004kalman}.
Therefore, the authors used a simplified simulation involving fixed observation points to experimentally approximate upper $\bar{\lambda}$ and lower $\underline{\lambda}$ heuristics for $\mb{\Lambda}$ of the IKD reward function $\mc{Q}$ \eqref{eq:ikd:reward:collab} as applied to Kalman filters for use in the full simulation.

\subsection{Model Formulation} \label{sec::exp::model}
Due to size, weight, and power (SWAP) restrictions of the UAVs, they are assumed to have heterogeneous sensors with overlapping sensor support.
Each UAV has been assigned to carry two sensors on their platform from a total possibility of three:
\begin{enumerate*}
  \item RF geolocation,
  \item optical tracking, and
  \item laser range finder.
\end{enumerate*}
Therefore each node has the option of sending the result of a sensor to any one of the nodes that it's connected to.
This creates an action space that as previously described combines the sensors' data on-board and the number of neighbors $\mc{N}_i$, including the action of not transmitting any information.

\begin{remark}
  The combinatorial explosion caused by scaling the number of agents in the system can be combatted with the use of collaboration graphs to factor the action space between agents by clustering capabilities, location, etc.
\end{remark}

As an example, a UAV connected to two other drones with RF geolocation and optical tracking capabilities, there are five total actions to choose during any time epoch:
\begin{enumerate}
  \item Do not communicate (Silence or Null)
  \item Send RF Geolocation results to Node A
  \item Send Optical Tracking results to Node A
  \item Send RF Geolocation results to Node B
  \item Send Optical Tracking results to Node B
\end{enumerate}

The state information is assumed to be a combinatorial set of the relevance of the data, the risk to the ground asset, and the state of collaboration based on the expected value of information from previous communications and the quality of the data.
The rewards are also combinatorial as formulated in IKD with the information theoretic metric $\rho$ based upon the information matrix of the sensor variances $\rho=1/\kappa_a$.
This leads to an updated relevance of
\begin{align}
  R_r (a,s) &= \mc{L}(s' \notin \mb{F} | a, s ) \cdot \frac{1}{\kappa_a},
\end{align}
%
where $\kappa_a = \min_{j \in \mc{N}} (\sigma_i \sigma_j)$,
$\sigma_i$ is the variance of the local sensor information being transmitted, and
$\sigma_j$ is the variance of a collaborative agent $j \neq i \in \mc{N}$.
This also leads to an updated collaboration reward of
\begin{align}
  \hat{R}_c (a,s) &= \frac{1}{|\mc{N}|} \sum_{j \in \mc{N}} \mc{S}( \mc{Q}(\lambda_j|a,s) ) \cdot \frac{1}{\kappa_a},
\end{align}
where the sigmoid function $\mc{S}(\cdot)$ has a center point $x_0 = \frac{1}{2} (\bar{\lambda} + \underline{\lambda})$ and k is set to create a switching function beyond the heuristic bounds of $\mb{\Lambda}$.
The function $\mc{Q}$ adjusts $\lambda_j$ based upon the current state $s$ of communication collaboration and the value of information $\kappa_a$ in the action $a$ to improve the covariance of the Kalman filter.

The transition probabilities for relevance are driven by the relationship between the monitoring UAVs and the ground asset.
The ground asset is assumed to be reachable by the UAVs through the MANET or a multi-hop network so that it can be influenced in finding a more conservative route by updating the edge costs between waypoints around the relative position of the hazard.
The timeliness and usefulness of the data being transmitted and received was utilized to calculate the collaboration transition probabilities via repeated simulation.

Observation probabilities are defined through subject matter expertise and repeated simulation.
The Kalman filter and a sufficient statistic of observation histories, a Bayesian estimator, were processed through a belief function $\upsilon$ to create the \emph{relevance} and \emph{collaboration} mapping to the observation space $o \in \mb{O}$.
The observation states themselves are a discrete categorization of local observations using a K-means clustering algorithm learned unsupervised from numerous Monte Carlo simulations.
In particular, the mapping of observations $\upsilon_r$ is performed solely by cluster classification of the state estimation and its covariance, driven by sensor inaccuracies of the ground asset's proximity to hazards to observation relevance states $o_r$.
Whereas, the observation mapping $\upsilon_c$ for collaboration is based on multiple observations:
\begin{enumerate*}
    \item a Bayes estimation of the current arrival probability $\hat{\lambda}_{ij}$ of messages $\sigma: j \rightarrow i$ or $\sigma: i \rightarrow j$ used to ascertain proximity to heuristic bounds;

    \item the usefulness of information $\sigma_k$ recently received from or sent to another agent $j$ according to $\kappa_a$; and,

    \item the lagged autocovariance $cov(\mb{P}[m], \mb{P}[m+l])$ of the Kalman filter covariances $\mb{P}$ to evaluate the impact of information exchange over time in maintaining mission objectives involved in asset tracking.
\end{enumerate*}
A \emph{collaboration} mapping $\upsilon_c$ is the cluster classification of $\hat{\lambda}$, $\sigma_k$, and $cov(\cdot)$ to the observation space $o_c$.
The relevance $o_r$  and collaboration $o_c$ observations from the mappings $\upsilon$ are indexed to the observation $o$ in the combinatorial set, which then drives the edge transitions of the finite state controller policy.

\subsection{Constraints on Communication}
The constraints being analyzed by CA-POMDP are the utilization of bandwidth (bytes per second) and power (Watt per second) over one second, which is a sampling length of 10 epochs as the UAV makes a collaboration decision every $100ms$.
The constraints are considered soft constraints because there is often more available bandwidth in the network than is being allocated to multi-agent collaboration.
Also, network protocols like ZeroMQ permit queueing techniques to account for myopic link saturation.
The power utilization is considered an average power consumption needed for the communication system to use to ensure the UAV can maintain the required flight time needed to maintain situational awareness of the disaster site.

Resource utilization by the UAVs is dependent on the type of information they are sending.
It is assumed that sending optical data requires more data and therefore higher power requirements to achieve the SNR needed for higher bandwidth physical layer protocols, e.g., 64QAM.
The data for laser range finding then RF geolocation are assumed to take sequentially less data and power than optical data.

\section{Experimental Results \& Analysis} \label{sec::results}
The applicability of Constrained-Action POMDPs to address resource awareness in Intelligent Knowledge Distribution was investigated through a series of Monte Carlo simulations.
From these simulations, the probabilistic constraint satisfaction was analyzed between intelligent (IKD), greedy, and naive communication models while confirming their ability to maintain acceptable state estimations.

\begin{figure}[!t]
  \centering
  \begin{subfigure}[b]{\columnwidth}
    \centering
    \includegraphics[width=0.75\textwidth]{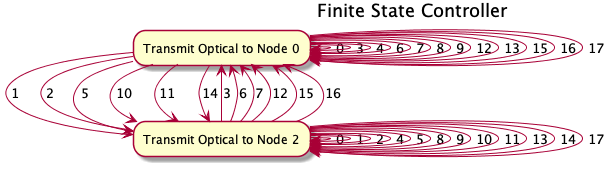}
    \caption{\footnotesize An unconstrained optimal FSC policy.}
    \label{fig:results:controller:unc}
  \end{subfigure}
  \begin{subfigure}[b]{\columnwidth}
    \includegraphics[width=\textwidth]{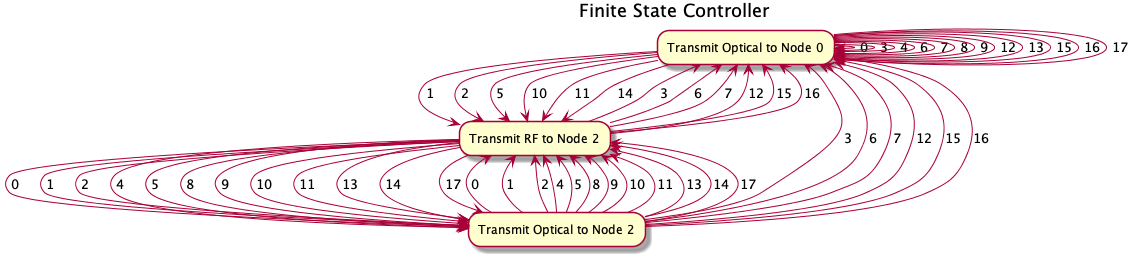}
    \caption{\footnotesize A constrained FSC policy where a less optimal but less resource consuming action (Transmit RF) was introduced into the controller.}
    \label{fig:results:controller:con}
  \end{subfigure}
  \caption{\footnotesize An example of (a) an unconstrained and (b) an constrained controller of one of the drones (agents) performing Intelligent Communications.}
  \label{fig:results:controller}
\end{figure}

\begin{figure}[!t]
    \centering
  \includegraphics[width=\columnwidth]{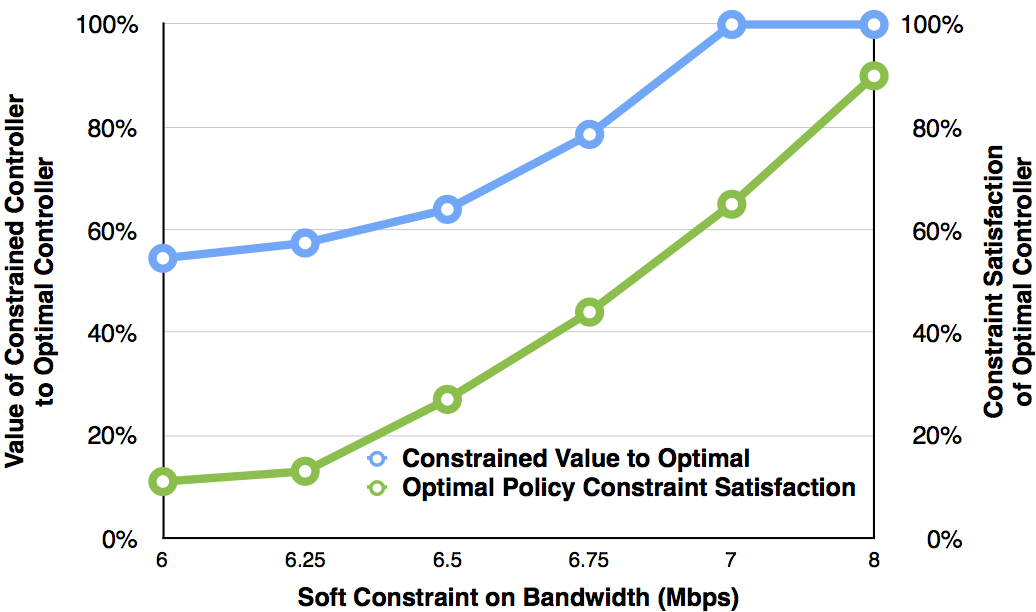}
  \caption{\footnotesize The impact to the value function and probabilistic constraint satisfaction when satisfying soft limits for Bandwidth. The green line shows the probabilistic constraint satisfaction of an Optimal FSC policy that only satisfies 12\% at 6 MBps. The blue line shows the value impact of a Constrained Controller that is 55\% of optimal at 6 MBps.}
  \label{fig:results:value}
\end{figure}

In our experiment, greedy communications assume a constant communication of the best information observed at all times alternating the destination between each time epoch among collaborating neighbors.
Naive communications applies a probability of communicating depending on the relevance of the information, the estimated proximity to a hazard.
In previous work \cite{fowler2018icra}, we already examined the trade off of constraint satisfaction and optimal value, Figure~\ref{fig:results:value}, along with an analysis of how the finite state controller is impacted by constraint states, Figure~\ref{fig:results:controller}.

The probabilistic constraint satisfaction between the three communication models was analyzed in a Monte Carlo simulation where the simulation would randomly determine a start location for the ground vehicle, a goal location for the ground vehicle, placement of hazards in the environment, and an initial belief state for the controller.
The UAV monitoring drones had a predefined search pattern that was consistent between all simulations.
During the IKD simulations, the CA-POMDP model was allowed to recalculate a new constrained policy during the execution of a simulation when the variation of information exceeded a threshold to ensure continued compliance to constraints.

\begin{figure}[!t]
  \centering
  \begin{subfigure}[b]{\columnwidth}
  \centering
    \includegraphics[width=0.9\textwidth]{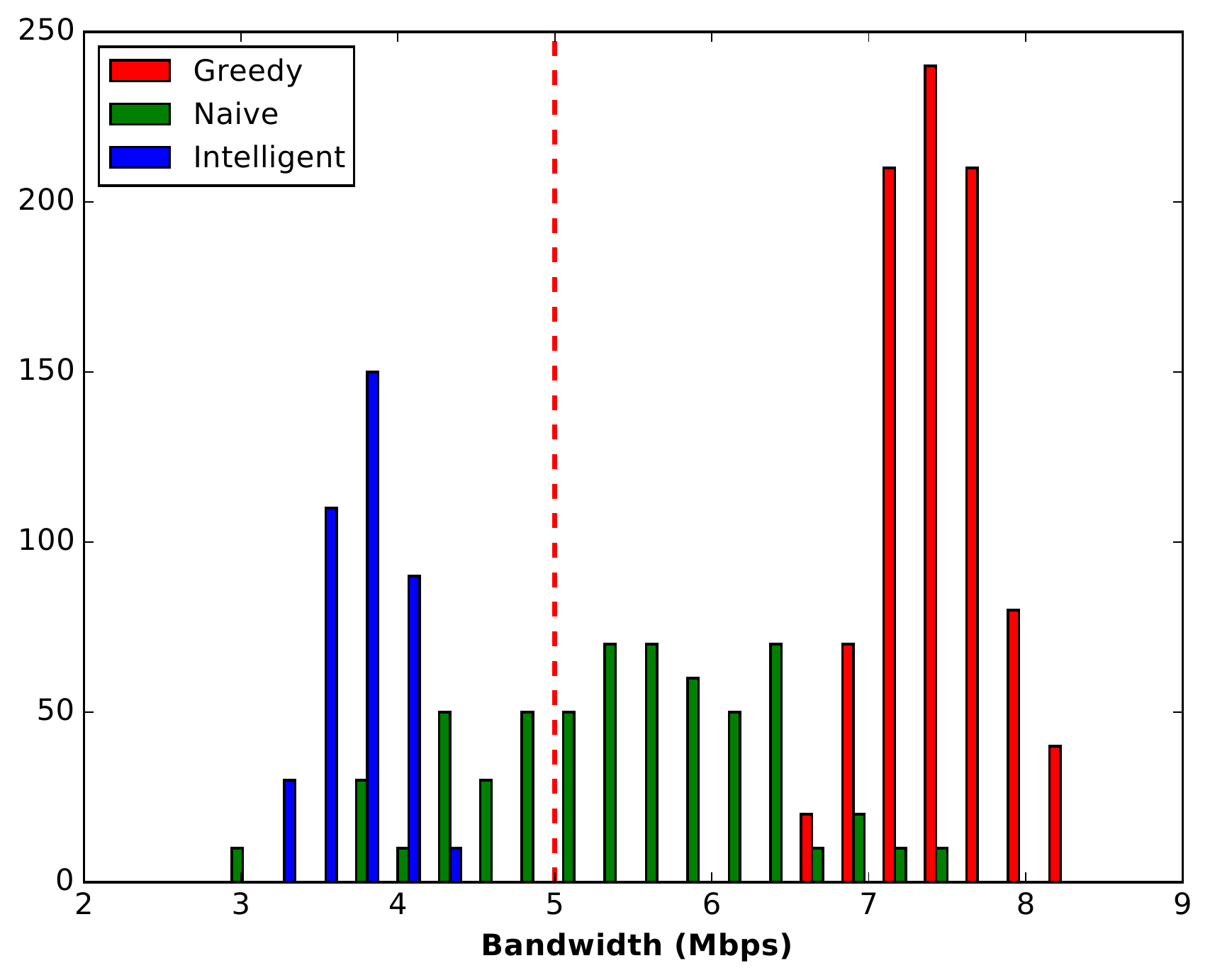}
    \caption{\footnotesize Bandwidth}
    \label{fig:results:constraints:bw}
  \end{subfigure}
  \begin{subfigure}[b]{\columnwidth}
    \centering
    \includegraphics[width=0.9\textwidth]{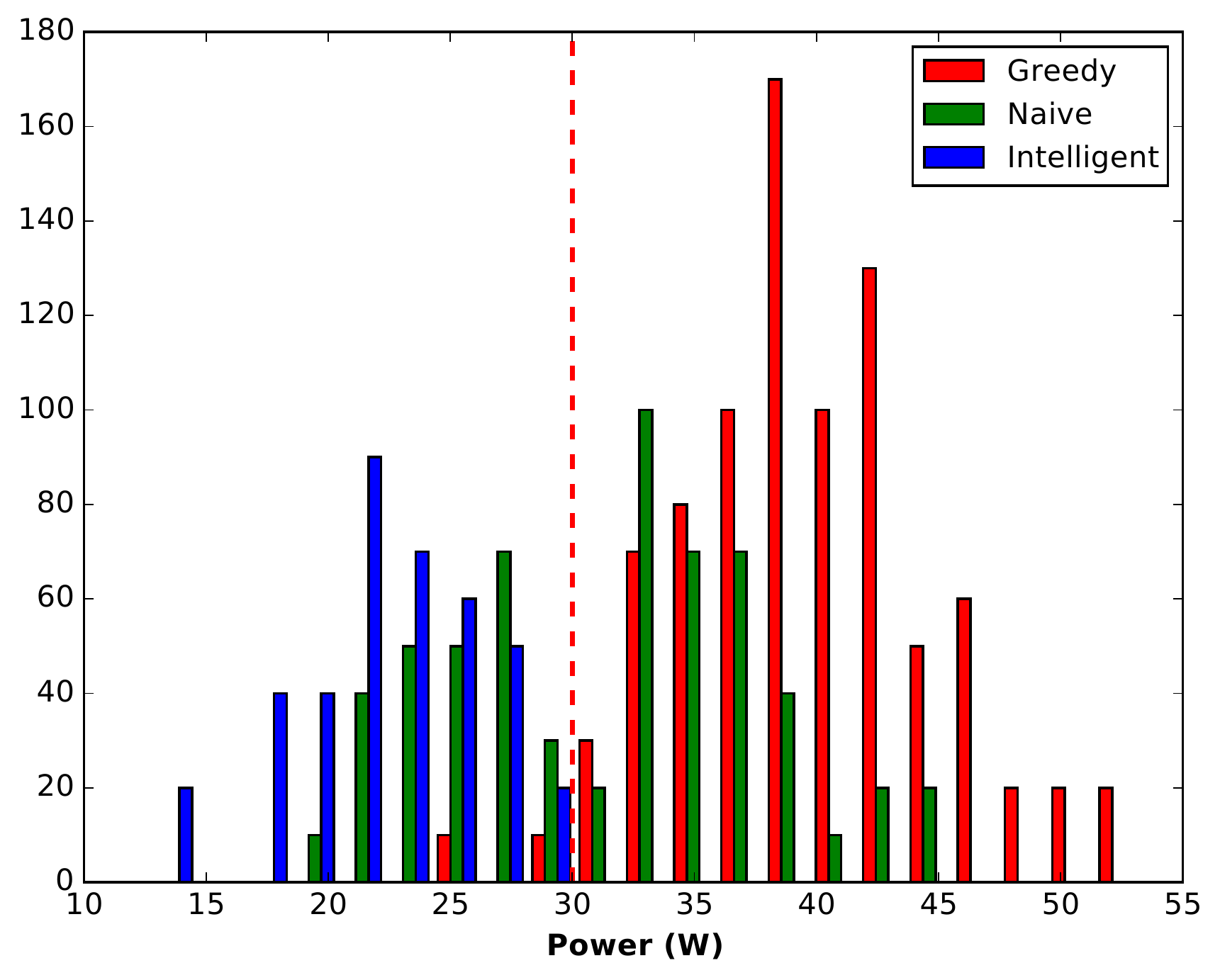}
    \caption{\footnotesize Power}
    \label{fig:results:constraints:pwr}
  \end{subfigure}
  \caption{\footnotesize Comparison of constraint satisfaction from three Monte Carlo simulation scenarios involving drones performing either Greedy, Naive, or Intelligent Communications.}
  \label{fig:results:constraints}
\end{figure}

The utilization of resources per second by each agent during all these simulations was graphed as histograms in Figure~\ref{fig:results:constraints} for the two limiting resources along with their soft constraints (dotted red line).
The probabilistic constraint satisfaction to these soft limits of both bandwidth and power were substantially improved for intelligent knowledge distribution compared to those of both naive and greedy communication models.
This confirms that the CA-POMDP substantiation for \emph{Intelligent Knowledge Distribution} was able to find a controller that complied with probabilistic constraints against soft limits and adapt the controller as needed.
During execution, the constraint state ``Transmit RF to Node 2'' of Figure~\ref{fig:results:controller:con} would be taken at appropriate times to reduce resource usage.

\begin{figure}[!t]
  \centering
  \includegraphics[width=0.9\columnwidth]{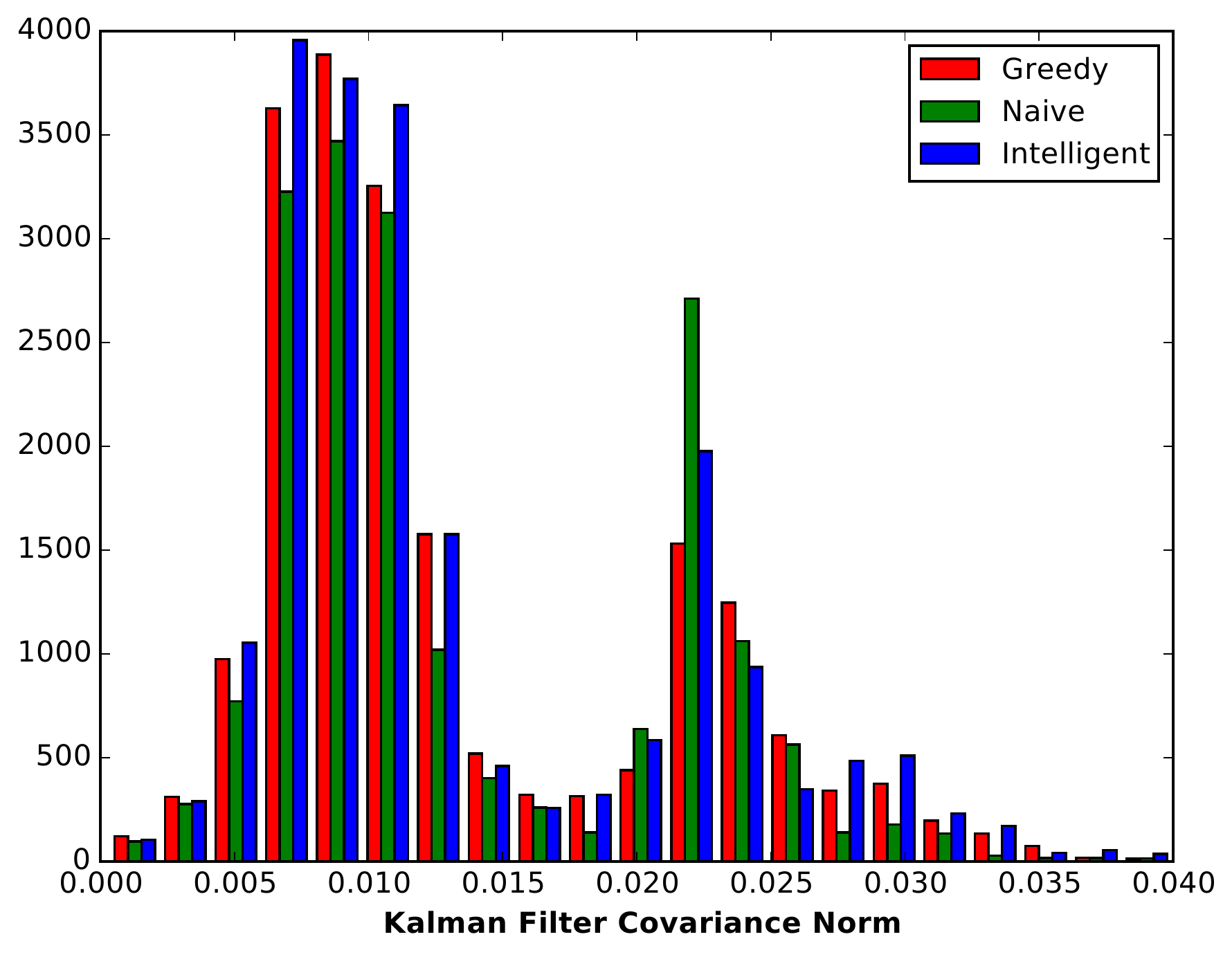}
  \caption{\footnotesize Histogram of the norm of the Unscented Kalman Filter covariance matrix showing that despite the constrained actions of IKD, the intelligent controller was still able to maintain accurate estimates of the ground vehicle.}
  \label{fig:results:accuracies}
\end{figure}

As expected, the naive communication model with its probabilistic approach to making communication decisions on data relevance alone utilized less resources than the purely greedy communication model.
Despite the improvement shown with the naive approach, our IKD approach was compliant to the soft constraints while not negatively impacting the state estimation, determined by the norm of the UKF covariance in Figure~\ref{fig:results:accuracies}.
The two distributions in the graph are indicative of the times when the filter has full observational data and when the filters have missing observations and measurements.

The first two columns of Table~\ref{table::results::comparison} further validates our results showing the probabilistic constraint satisfaction, or CDF below soft constraints, to the bandwidth and power limits.
The accuracy of the state estimations are shown in the last column and it is of interest to note that IKD maintained better estimations than naive.
It is hypothesized that the improvement results from IKD sending the most relevant data to ensure bounded performance of the Kalman filter, whereas naive does not have that level of contextual understanding.

\begin{table}[!ht]
  \centering
  \caption{Comparison of Simulation Metrics between Greedy, Naive, and Intelligent Communications.}
  \label{table::results::comparison}
  \begin{tabular}{lccc}
    \toprule
    \bf Model & \bf BW  & \bf Power & \bf Cov \\
    \midrule
    Greedy        & 0\% & 2\% & 0.0132 \\
    Naive         & 32\% & 40\% & 0.0135 \\
    Intelligent   & 97\% & 97\% & 0.0133 \\
    \bottomrule
  \end{tabular}
\end{table}

\begin{remark}
  It is interesting to note that during the construction and analysis of the system that if the optimal controller meets probabilistic constraint satisfaction less than 15\% of the time, then there was an unacceptable level of probability that the the resulting constrained controller would not meet mission objectives.
  Future work can investigate the driving mechanism behind mission success and techniques to compensate like adaptive soft constraints to ensure mission objectives.
\end{remark}

\section{Future Work \& Conclusion}
\label{sec::future}

The Monte Carlo simulations indicate that the incorporation of Intelligent Knowledge Distribution (IKD) that determines what information is needed to whom and when under soft constraints was successful.
The approach provides an \emph{agnostic} ``plug-and-play'' framework for IKD by constraining the actions of a POMDP to only transmit information to a collaborative agent when the value of that information warrants the communication.
Constrained-Action POMDPs provide a level of guarantee of probabilistic constraint satisfaction to a desired operational behavior while still allowing short-term bursts of critical information.
It also validates the concept of using Markov chain Monte Carlo analysis to evaluate an infinite-horizon policy represented as a finite state controller for its probability of satisfying soft constraints and a combinatorial discrete optimization for achieving desired constraint behavior of a policy while minimizing impact on the controller's value.

As the number of agents grow to respond to operational needs, communication between agents grows according to $\frac{1}{2}n(n-1)$.
To reduce the number of communication channels and improve the feasibility of solving for the IKD communication model in larger multi-agent systems, learning coordination graphs seen in Network Distributed POMDPs (ND-POMDP) implementations will be addressed in future work, which could be different for each taxonomy of information being tracked by the agents.
For IKD in real-time operational scenarios, the agents available to collaborate and coordinate may not be known until they are actively in the field, therefore learning heterogeneous interactions and constructing a coordination graph online is a necessary expansion.  Finally, the above approach is actively being expanded into a fully collaborative multi-agent target tracking simulation to evaluate its effectiveness in a Concurrent Constrained Decentralized POMDP, where the IKD CA-POMDP is one POMDP working concurrently with tasking, motion planning, and target tracking MDPs.

\appendices

\ifCLASSOPTIONcaptionsoff
  \newpage
\fi
\bibliographystyle{IEEEtran}
\bibliography{ms}

\begin{IEEEbiography}[{\includegraphics[width=1in,height=1.25in,clip,keepaspectratio]{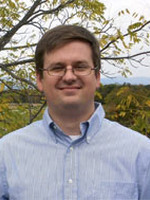}}]{Michael Fowler}
Michael Fowler is a research scientist for the Ted and Karyn Hume Center for National Security and Technology at Virginia Tech responsible for driving and providing thought leadership into the centers research on autonomous systems, mission orchestration, distributed intelligence, and security for wireless and unmanned systems with currently over \$4M in active research programs.
His research focus is on the convergence of distributed intelligence and machine learning for decision making under uncertainty for embedded applications including drones, satellites, IoT, and wireless communications.
He has accumulated over 10 years of experience managing and performing research in security, wireless systems and artificial intelligence at Harris Corporation and as faculty at Virginia Tech.
He has received a Masters in Engineering Management from Old Dominion University and is currently a Ph.D. candidate in Computer Engineering at Virginia Tech.
\end{IEEEbiography}

\begin{IEEEbiography}[{\includegraphics[width=1in,height=1.25in,clip,keepaspectratio]{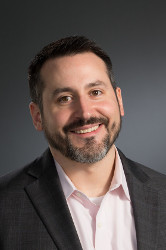}}]{T. Charles Clancy}
Dr. Charles Clancy is an Associate Professor of Electrical and Computer Engineering at Virginia Tech and directs of the Hume Center for National Security and Technology.
Prior to joining Virginia Tech in 2010, he served as a senior researcher at the Laboratory for Telecommunications Sciences, a defense research lab at the University of Maryland, where he led research programs in software-defined and cognitive radio.
Dr. Clancy received his B.S. in Computer Engineering from the Rose-Hulman Institute of Technology, M.S. in Electrical Engineering from the University of Illinois, and his Ph.D. in Computer Science from the University of Maryland.
He is a Senior Member of the IEEE and has over 150 peer-reviewed technical publications.
His current research interests include cognitive communications and spectrum security.
\end{IEEEbiography}

\begin{IEEEbiography}[{\includegraphics[width=1in,height=1.25in,clip,keepaspectratio]{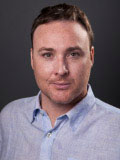}}]{Ryan Williams}
Ryan K. Williams received the B.S. degree in computer engineering from Virginia Polytechnic Institute and State University in 2005, and the Ph.D. degree from the University of Southern California in 2014.  He is currently an Assistant Professor in the Electrical and Computer Engineering Department at Virginia Polytechnic Institute and State University where he runs the Virginia Tech Laboratory for Coordination at Scale (CAS Lab).  His current research interests include control, cooperation, and intelligence in distributed multi-agent systems, topological methods in cooperative phenomena, and distributed algorithms for optimization, estimation, inference, and learning.  Williams is a Viterbi Fellowship recipient, has been awarded the NSF CISE Research Initiation Initiative grant for young investigators, is a best multi-robot paper finalist at the 2017 IEEE International Conference on Robotics and Automation, and has been featured by various news outlets, including the L.A. Times.
\end{IEEEbiography}

\end{document}